\documentclass[10pt, letterpaper]{article}
\usepackage[top = 0.85in, left = 2.25in, footskip = 0.75in]{geometry} 
\usepackage[shortlabels]{enumitem} 
\usepackage{amsmath, amsthm, amssymb}
\usepackage{mathtools} 
\usepackage{changepage}

\usepackage{textcomp, marvosym, adjustbox}

\usepackage{cite}

\usepackage{nameref, hyperref}

\usepackage[right]{lineno}

\usepackage[nopatch = eqnum]{microtype}
\DisableLigatures[f]{encoding = *, family = *}

\usepackage[table]{xcolor}

\usepackage{array}

\newcolumntype{+}{!{\vrule width 2pt}}

\newlength\savedwidth

\usepackage{ragged2e} 
\justifying 
\usepackage[aboveskip = 2pt, labelfont = bf, labelsep = period, singlelinecheck = off]{caption}

\makeatletter
\renewcommand{\@biblabel}[1]{\quad#1.}
\makeatother

\usepackage{lastpage, fancyhdr, graphicx}
\usepackage{epstopdf}

\fancyheadoffset[L]{2.25in}
\fancyfootoffset[L]{1.75in} 
\begin{document}
\vspace*{0.125in}


\begin{flushleft}

{\Large
    \textbf\newline{Initial Luminally Deposited FGF4 Critically Influences Blastocyst Patterning}
}
\newline
\\
Michael Alexander Ramirez Sierra\textsuperscript{1,2,*},
Thomas R. Sokolowski\textsuperscript{1}
\\
\bigskip
\textbf{1} Frankfurt Institute for Advanced Studies (FIAS), Ruth-Moufang-Straße 1, 60438 Frankfurt am Main, Germany
\\
\textbf{2} Goethe-Universität Frankfurt am Main, Faculty of Computer Science and Mathematics, Robert-Mayer-Straße 10, 60054 Frankfurt am Main, Germany
\\
\bigskip

* ramirez-sierra@fias.uni-frankfurt.de

\end{flushleft}

\graphicspath{{./Manuscript_Graphics/}} 



\section*{Abstract}

Luminogenesis, the formation of a fluid-filled cavity (lumen), is an essential process in early mammalian embryonic development, coinciding with the second cell-fate decision that differentiates the inner-cell-mass (ICM) into epiblast (EPI) and primitive endoderm (PRE) tissues. Based on experiments, the blastocyst lumen is hypothesized to influence EPI-PRE tissue specification, but its particular functional role remains theoretically underexplored. In this study, we extended our stochastic ICM differentiation model to incorporate both the blastocyst lumen (blastocoel) and the trophectoderm (TE) as adjacent compartments where the primary signaling protein (FGF4) for EPI-PRE differentiation can diffuse, degrade, or accumulate. This extended ICM model allows for a spatially resolved analysis of EPI-PRE lineage proportioning under the influence of luminally deposited FGF4 molecules. Our results reveal that the blastocoel acts as a localized signaling source, while the TE functions as an embryo-wide signaling sink, guiding cell-fate decisions within the ICM. A critical determinant of the ideal target system behavior is the initial amount of luminally deposited FGF4, which is required to recapitulate the correct spatio-temporal patterning of EPI and PRE (blastocyst) cell lineages. Notably, this requirement is independent of ICM population size and shape, highlighting the robustness of the FGF4 signaling process. Our study also underscores the potential of integrating single-cell gene expression and cell-cell communication dynamics simulations with tissue-level morphogenesis representations. By combining spatial-stochastic modeling with agent-based frameworks, we could enhance the exploration of the intricate interplay between gene regulation, signaling, and morphogenetic processes that govern early embryonic development. Understanding these mechanisms is fundamental for deciphering the principles underlying successful blastocyst patterning and may have broad implications for studying human early development.


\section*{Author summary}

We present an extension to our ICM model that includes an explicit notion of the blastocyst lumen (blastocoel), a fluid-filled cavity, and the trophectoderm (TE), a protective tissue which constitutes the embryonic part of the placenta. We extended our spatial-stochastic simulator by incorporating the blastocoel and the TE as separate (static) compartments adjacent to the ICM, and by exploiting the AI-powered SBI workflow built in \cite{ramirez-sierra_ai-powered_2024} and \cite{ramirez_sierra_comparing_2025}, we parameterized this extended ICM model. On this basis, we studied the synergistic interplay between intracellular biochemical noise and intercellular communication, while considering the effects of the presence of blastocoel and TE, for achieving the correct blastocyst patterning. Here, the key signaling molecule for EPI-PRE differentiation (FGF4) can only be produced by the ICM population, and the two adjacent compartments act as FGF4 reservoirs where it can either degrade or continue to diffuse across blastocyst constituents almost freely. Our simulations indicate that, when there is a substantial amount of initial luminally deposited FGF4 proteins, our extended ICM model perfectly recapitulates the ideal target system behavior: the ICM layer in contact with the blastocoel exhibited exclusive commitment to the PRE fate, whereas the remaining ICM layer(s) exhibited exclusive commitment to the EPI fate. Moreover, this ideal target system behavior shows signatures of strong reproducibility, as well as moderate robustness to perturbations of initial luminally deposited FGF4, highlighting the critical roles of the blastocoel and the TE in attaining the correct blastocyst (spatial) pattern.


\section*{Motivation and Background}

Luminogenesis, the formation of a fluid-filled cavity (lumen), is an elementary process of morphogenesis \cite{dumortier_hydraulic_2019, ryan_lumen_2019, chan_integration_2020, shahbazi_mechanisms_2020, kim_deciphering_2021, yanagida_cell_2022, le_verge-serandour_blastocoel_2022, fuji_computational_2022}. During mammalian preimplantation development, after the first cellular fate decision gives rise to trophectoderm (TE) and inner-cell-mass (ICM) tissues, the formation of the blastocyst lumen (also known as blastocoel) and the second cellular fate decision begin simultaneously \cite{ryan_lumen_2019}. Although this second cellular fate decision, which gives rise to epiblast (EPI) and primitive endoderm (PRE) cell populations, necessitates mechanical cues for correct cellular positioning and self-organized ICM tissue remodeling \cite{dumortier_hydraulic_2019, ryan_lumen_2019, shahbazi_mechanisms_2020}, only a few studies have properly investigated the functional roles played by the TE and, especially, the blastocyst lumen \cite{dumortier_hydraulic_2019, ryan_lumen_2019, kim_deciphering_2021}.

The TE forms an epithelium (a protective cell layer) that supports the interaction between maternal and embryonic tissues, mechanically constraining the ICM and constituting the embryonic part of the placenta after implantation \cite{plusa_common_2020, chowdhary_journey_2022}. Similarly, it is hypothesized that the blastocyst lumen not only mechanically induces indirect cellular responses to adjust embryonic tissue properties, but also directly acts as a biochemical signaling niche to control embryonic tissue patterning \cite{chan_integration_2020, shahbazi_mechanisms_2020}. The experimental work \cite{ryan_lumen_2019} examined blastocoel formation with respect to EPI-PRE cell-lineage specification and spatial segregation. Their findings suggest that luminally deposited fibroblast growth factor 4 (FGF4) molecules mediate EPI-PRE cell-fate differentiation by collectively acting as a localized signaling source. Importantly, they discovered that there is extensive embryo-wide cytoplasmic-vesicle secretion around embryonic day (E) 3.0, whereby some of these vesicles contain FGF4. In particular, their results showed that luminally inhibited FGF4 signaling negatively impacts (impairs) EPI-PRE specification and sorting, while luminally enhanced FGF4 signaling positively impacts (expedites) both of these processes. For this purpose, inhibition of FGF4 signaling in the lumen is achieved by luminal deposition of an allosteric FGFR1 (the main receptor of FGF4) inhibitor, and enhancement of FGF4 signaling in the lumen is achieved by luminal deposition of exogenous FGF4 protein \cite{ryan_lumen_2019}. Altogether, the mouse experimental work \cite{ryan_lumen_2019} hints at an interplay between mechanical constraining and cell-position-specific induction of fate-determining gene expression for successful ICM differentiation. They provide experimental evidence that mechanical constraining is enabled by differential affinity of cell membranes to luminal cues, whereas cell-position-specific induction of fate-determining gene expression is guided by FGF4 signaling proteins deposited in the blastocoel.

The premises presented by the experimental study \cite{ryan_lumen_2019} are particularly interesting because they challenge the classical interpretations of EPI-PRE specification and sorting, where intercellular communication orchestrates fate specification \cite{shahbazi_mechanisms_2020, raina_cell-cell_2021} and cell-cell membrane-interface affinity dictates fate sorting \cite{yanagida_cell_2022, maitre_asymmetric_2016}. However, at least for EPI-PRE specification, this paradigmatic cell-fate proportioning process is robust and reproducible under highly variable (intrinsic) biochemical noise and (extrinsic) environmental conditions, even when the ICM progenitor population is isolated from the TE and the blastocoel \cite{saiz_growth-factor-mediated_2020, fischer_salt-and-pepper_2023}. The presence of robustness and reproducibility despite vastly distinct biophysical factors highlights the critical crosstalk between (cell-level) gene regulation and (tissue-level) morphological change necessary for the correct progression of developmental biology systems.

In this study, by exploiting the computational tools established in \cite{ramirez-sierra_ai-powered_2024} and \cite{ramirez_sierra_comparing_2025}, we explore an extension of the Inferred-Theoretical Wild-Type (ITWT) system, where we not only incorporate an abstraction of the blastocyst lumen but also integrate the TE as an ``embryo-wide signaling sink''. We find that our extended ITWT model perfectly recapitulates the ideal target system behavior: the single lower cell layer (which is in contact with the blastocoel) exhibited exclusive commitment to the PRE fate, whereas the (multiple) upper cell layer(s) exhibited exclusive commitment to the EPI fate. This ideal target behavior holds irrespective of system size (number of ICM cells) and shape (number of cell layers). We remark that our extended model also inherits two fundamental characteristics of the original ITWT system: (1) the target behavior develops within 48 hours between E2.5 and E4.5; (2) the ICM adopts almost exclusively the EPI fate in the absence of any FGF4 signaling, and all the ICM cells start with the undifferentiated (UND) fate.

Overall, we observe that the blastocyst lumen and the TE help canalize and maintain embryonic patterning. However, the ideal target behavior only emerges when the system has a substantial amount of initial luminally deposited FGF4 molecules, which we characterize and quantify in the following.


\section*{Results}

\subsection*{Model parameter posterior distributions perfectly recapitulate ideal target system behavior}

To test the effects of the presence of blastocoel and TE on the establishment and stability of the correct blastocyst patterning, we incorporated them into the original ITWT system (see \cite{ramirez-sierra_ai-powered_2024}) as separate (static) components adjacent to the ICM. Therefore, as the ICM population produces the key signaling molecule (FGF4) for EPI-PRE specification, the two adjacent compartments (blastocoel and TE) act as reservoirs where FGF4 can either degrade or continue to diffuse (jump) across blastocyst constituents almost freely, which adds another dimension of complexity to the modeling; see Methods section for details. We also modified the original ICM architecture: instead of mirroring a simple rectangular monolayer of cells, the extended ITWT system uses a 2D projection of a pyramid-like structure which resembles the geometry of an actual blastocyst ICM. On the technical side, this abstracted ICM architecture facilitates the implementation of diverse simulation scenarios with not only distinct numbers of ICM cells and cell layers, but also seemingly arbitrary cellular arrangements, which mimics the irregular nature of ICM neighborhoods.

Accordingly, we performed eight independent model parameter inference runs of the AI-MAPE workflow (\cite{ramirez-sierra_ai-powered_2024} and \cite{ramirez_sierra_comparing_2025}), where each run consisted of five rounds (100 thousand simulations per round). Each independent inference run corresponds to a distinct system size-shape pair (number of ICM cells and number of cell layers), ranging from 2 ICM cells and 2 cell layers (1, 1) to 25 ICM cells and 5 cell layers (9, 7, 5, 3, 1); see Fig~\ref{ChapD_Fig1}. The first tuple entry represents the lowermost cell layer, whereas the last tuple entry represents the uppermost cell layer. We chose (5, 3, 1) as our ``representative reference'': 9 ICM cells and 3 cell layers (see cyan box of Fig~\ref{ChapD_Fig1}).

We obtained eight distinct model parameter distributions (posteriors) for the extended ITWT system; one posterior distribution for each of the respective (eight) ICM geometries. As seen in Fig~\ref{ChapD_Fig2}[A], the extension inherits the same original ITWT model basis; however, the extension also comprises several extra parameters necessary for modeling the transport/release of FGF4 across the blastocyst constituents: ICM (EPI plus PRE), blastocoel, and TE. As seen in Fig~\ref{ChapD_Fig2}[B], the blastocoel and the TE should have complementary functional roles (see the Methods section for additional details): the blastocoel should act as a localized signaling source, from which the initial luminally deposited FGF4 diffuses in the direction of the TE or the PRE-progenitor population (orange cell layer); the TE should act as an embryo-wide signaling sink, which accumulates FGF4 from the blastocoel or the ICM. Likewise, FGF4 diffuses (jumps) between neighboring constituents, and it can be degraded anywhere. We remark that the EPI-progenitor population (blue cell layer(s)) is uniquely capable of producing FGF4.

For the extended ITWT, the time-varying pattern score that quantitatively represents the system dynamics is a redefinition of the original ITWT objective function (see \cite{ramirez-sierra_ai-powered_2024}), which traces the deviation from the ideal EPI-PRE-UND lineage target proportions and positions: for the lower (orange) cell layer, 100\% PRE cells; for the upper (blue) cell layer(s), 100\% EPI cells. Similarly, the extended ITWT meta score $\overline{S}$, a single scalar that condenses the success of each MAPE (maximum a posteriori probability estimate) with respect to the target system behavior, is a redefinition of the original (see \cite{ramirez_sierra_comparing_2025}): $\overline{S} = \overline{\alpha}_{75} + \overline{\alpha}_{25} - \overline{\alpha}_{50}$. Here, $\overline{\alpha}_{25}$, $\overline{\alpha}_{50}$, and $\overline{\alpha}_{75}$ represent the time averages of the 25th, 50th, and 75th percentiles for an MAPE-associated \emph{de novo} simulation ensemble, which consists of 100 pattern score trajectories. As such, the ideal (or perfect) meta score is 1.

As shown in Fig~\ref{ChapD_Fig2}[C, D], all the posteriors (MAPEs) perfectly recapitulate the ideal target behavior, regardless of system size (number of ICM cells) and shape (number of cell layers). Moreover, we observe that the meta-score round progression towards the perfect state was rapid: by the second round, all the meta scores were above 0.7, and by the third round, all the meta scores were above 0.9. By the fifth round, all the meta scores reached 1, except for (9, 7, 5, 3, 1), whose meta score was just below 1 (0.98). These virtually ideal meta scores highlight the exceptional reproducibility of the correct blastocyst patterning exhibited by the extended ITWT system, considering the presence of blastocoel and TE and, more importantly, a significant amount of initial luminally deposited FGF4; this last aspect is crucial for perfectly recapitulating the ideal target system behavior, and it is explored in the following.

\begin{adjustwidth}{0in}{0in}
\includegraphics[width = 3.25in, height = 3.25in]{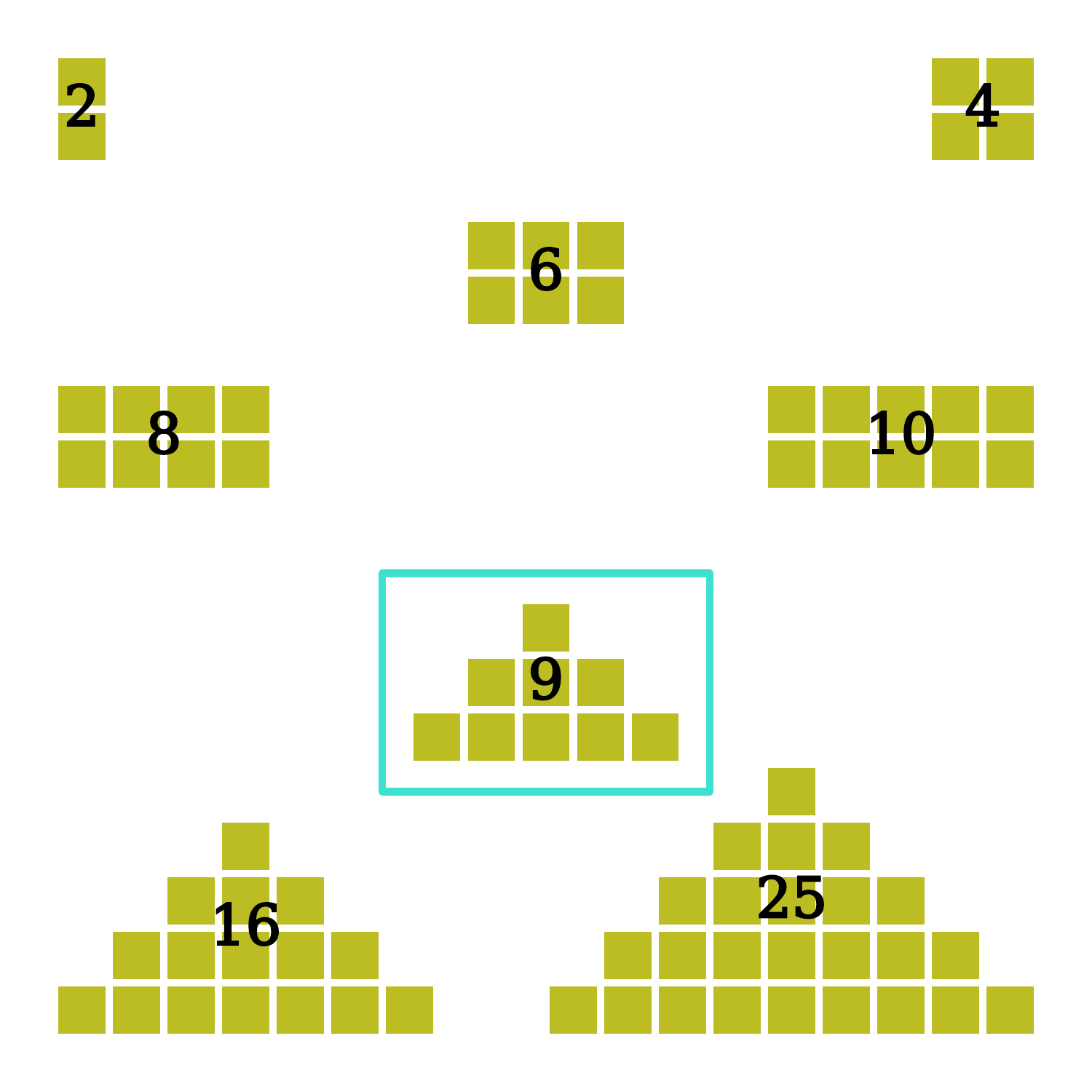} \centering 
\end{adjustwidth}
\begin{figure}[hpt!]
\caption{{\bf Overview of the extended ITWT system size-shape pairs.} We performed eight independent model parameter inference runs of the AI-MAPE workflow for the extended ITWT, where each run consisted of five rounds (100 thousand simulations per round). Each independent inference run corresponds to a distinct system size-shape pair (number of ICM cells and number of cell layers), ranging from 2 ICM cells and 2 cell layers (1, 1) to 25 ICM cells and 5 cell layers (9, 7, 5, 3, 1). We chose (5, 3, 1) as our ``representative reference'' (cyan box), corresponding to 9 ICM cells and 3 cell layers.}
\label{ChapD_Fig1}
\end{figure}

\subsection*{Comparison of all eight optimal parameter sets for the extended ITWT system}

In Fig~\ref{ChapD_Fig3}, we see the eight optimal parameter sets for the extended ITWT system obtained from the eight independent AI-MAPE workflow runs; one run per system size-shape pair. See also Table~\ref{ChapD_Table1} which summarizes model parameter definitions and explanations. The top-left panel shows the core gene regulatory network (GRN) interaction parameters, which dictate the cell-scale dynamics. The top-right and bottom-left panels show the spatial coupling parameters, which dictate the tissue-scale dynamics. The bottom-right panel shows the initial condition parameters: \textit{Nanog}-\textit{Gata6} mRNA, NANOG-GATA6 protein, and luminally deposited FGF4 protein copy numbers. Note that, for the extended ITWT, the most important additional parameters are $\tau_{\mathrm{LUMEN}}$, $\tau_{\mathrm{SINK}}$, and FGF4-LUMEN. Here, $\tau_{\mathrm{LUMEN}}$ is the mean FGF4 escape time from the blastocoel, and $\tau_{\mathrm{SINK}}$ is the mean FGF4 escape time from the TE.

Except for a few cases such as \textit{Gata6}\_NANOG, \textit{Fgf4}\_NANOG, \textit{Fgf4}\_GATA6, and $\tau_{\textrm{doh},\textrm{ERK}}$, where we notice weak agreement among parameter value sets, we observe fair agreement among all the parameter sets for predictions of the core GRN interaction values, as well as strong agreement among the predicted values for the spatial coupling and the initial condition parameters. This result is remarkable considering the high-dimensional objective parameter space. Surprisingly, we do not clearly identify any system size-shape dependency among the eight ITWT optimal parameter sets, even for the amount of initial luminally deposited FGF4, whose inferred parameter values fall within a rather constrained range.

We also find some interesting systematic parameter relationships. We observe that $\tau_{\mathrm{d},\mathrm{FGF4-SINK}}$ (TE-associated FGF4 lifetime) is consistently lower than $\tau_{\mathrm{d},\mathrm{FGF4-LUMEN}}$ (blastocoel-associated FGF4 lifetime), but $\tau_{\mathrm{SINK}}$ is consistently higher than $\tau_{\mathrm{LUMEN}}$. In the aggregate, these systematic relationships indicate that, indeed, relative to one another, the blastocoel acts as an FGF4 source and the TE acts as an FGF4 sink: the TE has a faster nominal FGF4 degradation rate but a slower nominal mean FGF4 escape rate than the blastocoel. Note that this predictive result is nontrivial because these two parameters share the same prior range, and there is no explicit inductive bias in the inference procedure.

To reinforce the notion of proximity among all the optimal parameter sets, we used the L1 metric (which is normalized between 0\% and 100\% with respect to the prior ranges) for quantifying distances between all the parameter vectors; see Fig~\ref{ChapD_Fig4}. Overall, we observe short distances (strong agreements) between all the optimal parameter sets.

\begin{adjustwidth}{0in}{0in}
\includegraphics[width = 5.5in, height = 5.5in]{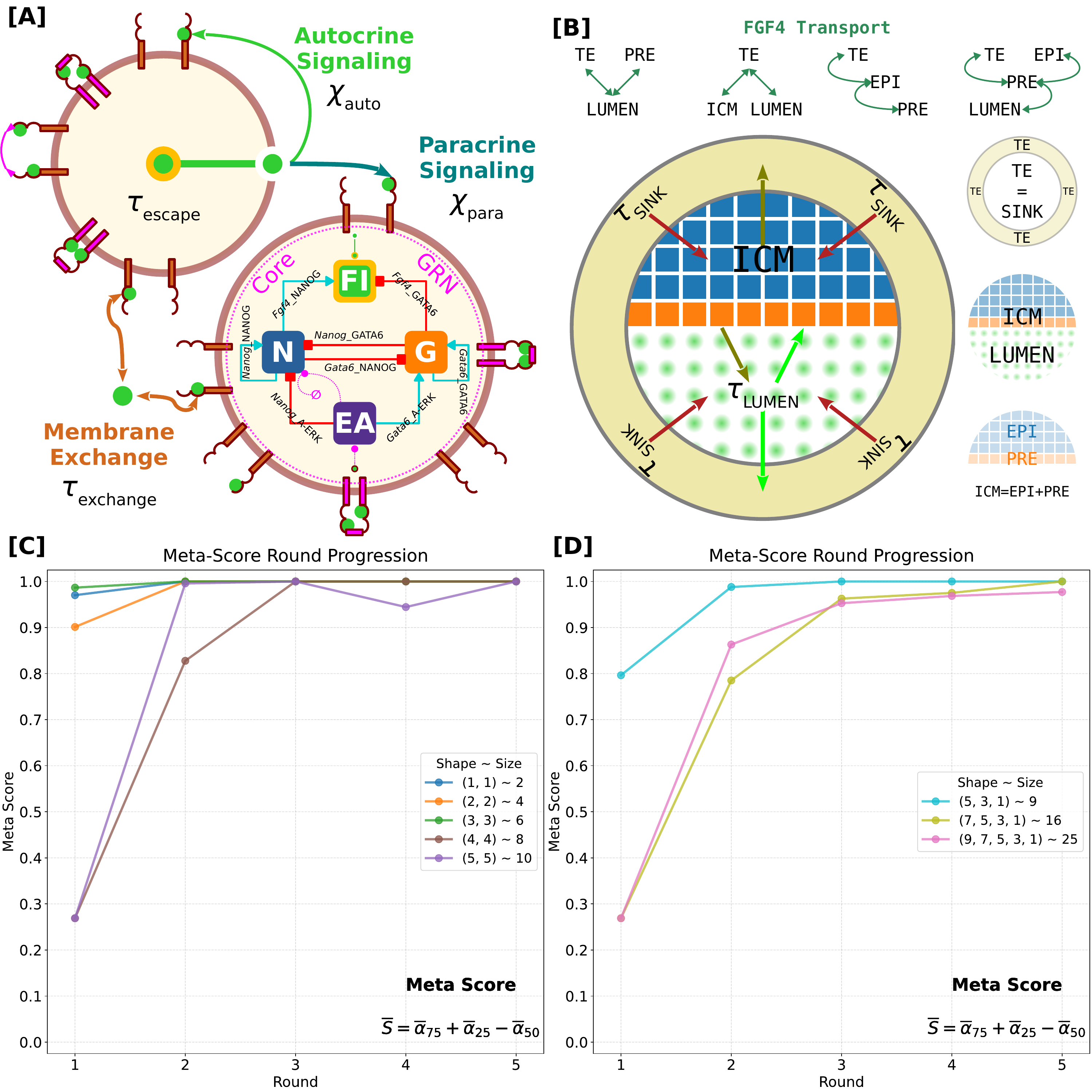} \centering 
\end{adjustwidth}
\begin{figure}[hpt!]
\caption{{\bf GRN motif and blastocyst-wide signaling model for the extended ITWT: parameter posterior distributions perfectly recapitulate the ideal target system behavior.} We obtained eight distinct model parameter distributions (posteriors) for the extended ITWT system. {\bf [A]} The extension inherits the original ITWT model basis (see \cite{ramirez-sierra_ai-powered_2024}). {\bf [B]} The extended ITWT model comprises several additional parameters necessary for modeling the transport/release of FGF4 across blastocyst constituents: ICM (EPI plus PRE), blastocoel, and TE. The blastocoel and the TE have complementary functional roles: the blastocoel acts as a localized signaling source, where the initial luminally deposited FGF4 diffuses in the direction of the TE or the PRE-progenitor population (orange cell layer); the TE acts as an embryo-wide signaling sink, receiving FGF4 from the blastocoel or the ICM. FGF4 diffuses (jumps) between neighboring constituents, and it can be degraded anywhere. The EPI-progenitor population (blue cell layer(s)) is uniquely capable of producing FGF4. {\bf [C, D]} All the posteriors (MAPEs) perfectly recapitulate the ideal target behavior, regardless of system size (number of ICM cells) and shape (number of cell layers).}
\label{ChapD_Fig2}
\end{figure}

\begin{adjustwidth}{0in}{0in}
\includegraphics[width = 5.9in, height = 5.9in]{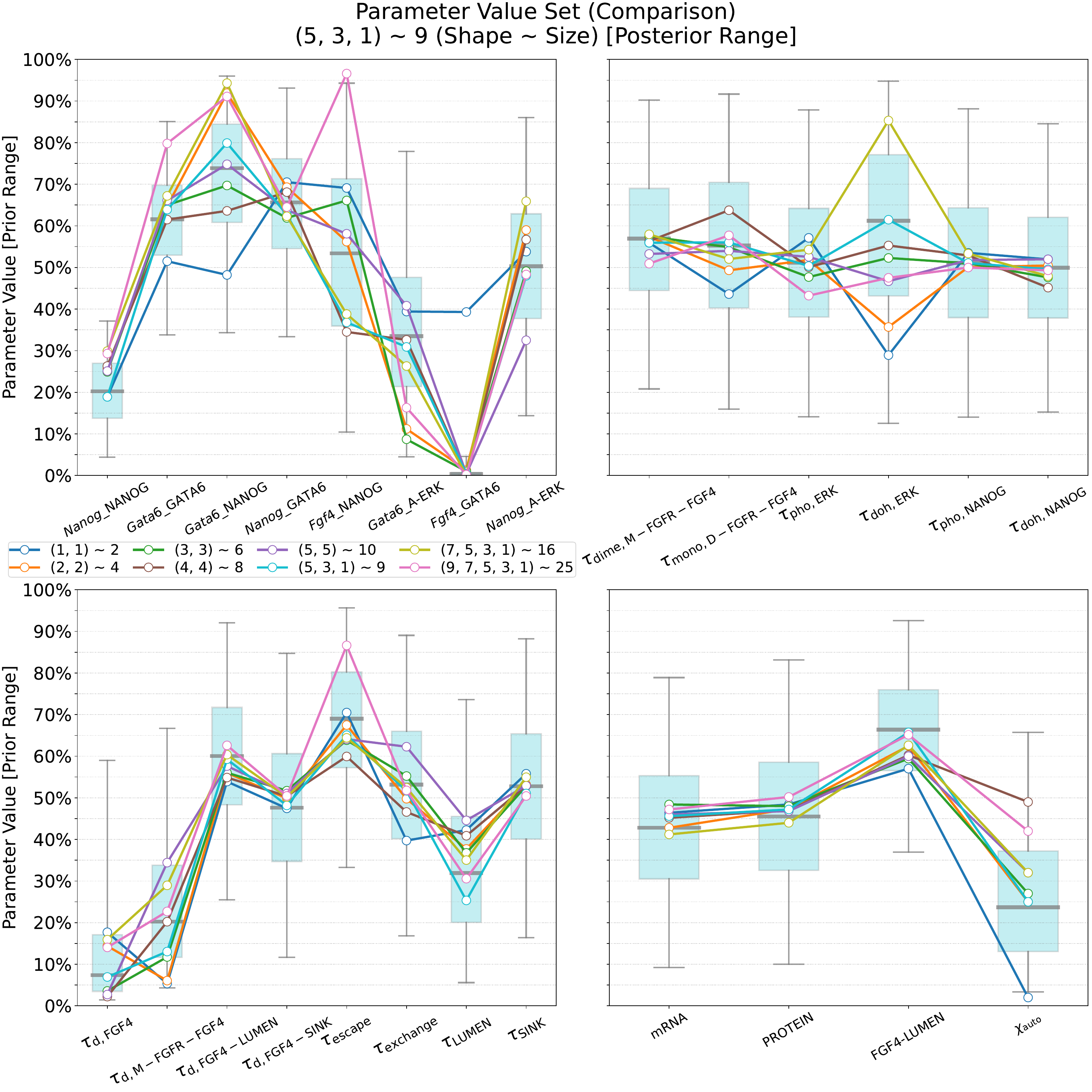} \centering 
\end{adjustwidth}
\begin{figure}[hpt!]
\caption{{\bf Comparison of all eight optimal parameter sets for the extended ITWT system.} Box-and-whisker diagrams for each (5, 3, 2, 1) one-dimensional marginal posterior distribution are shown as baselines. Boxes cover interquartile ranges (from the 25th to the 75th percentiles) and show medians (50th percentiles). Whiskers cover ranges from the 2.5th to the 97.5th percentiles. Parameter values fall within normalized prior ranges. The top-left panel shows the core gene regulatory network (GRN) interaction parameters, which dictate the cell-scale dynamics. The top-right and bottom-left panels show the spatial coupling parameters, which dictate the tissue-scale dynamics. The bottom-right panel shows the initial condition parameters. Value differences might indicate the emergence of potential compensation mechanisms among parameters, highlighting distinct exploitable strategies to achieve the ideal or target system behavior. Colors represent distinct system size-shape pairs: blue (1, 1); orange (2, 2); green (3, 3); brown (4, 4); purple (5, 5); cyan (5, 3, 1); olive (7, 5, 3, 1); pink (9, 7, 5, 3, 1).}
\label{ChapD_Fig3}
\end{figure}
\clearpage

\begin{adjustwidth}{0in}{0in}
\includegraphics[width = 5.75in, height = 5.75in]{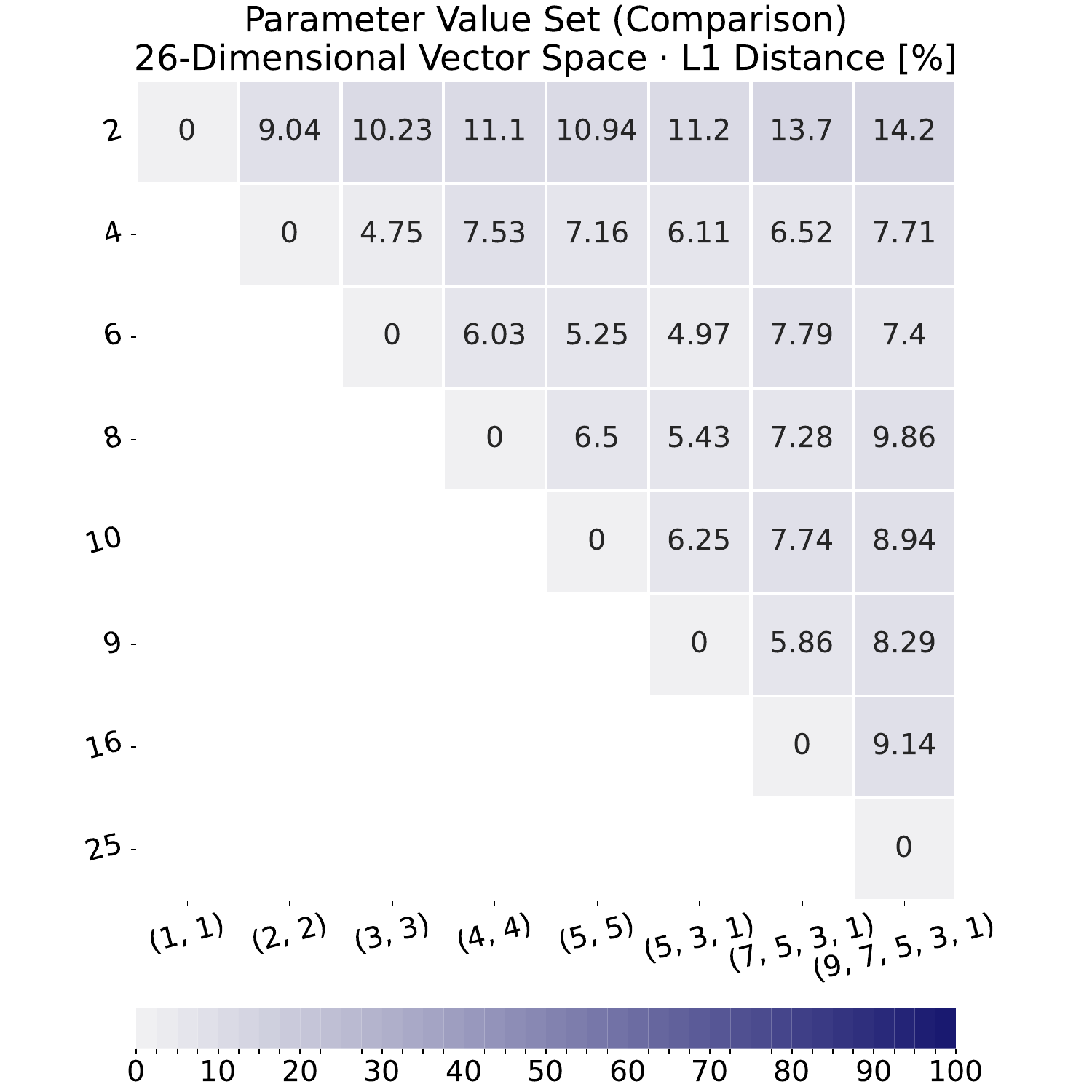} \centering 
\end{adjustwidth}
\begin{figure}[hpt!]
\caption{{\bf Distances between system size-shape pairs.} Distance matrix contrasting each pair of parameter sets: the parameter value sets are assumed to be elements of an abstract twenty-six-dimensional vector space. The L1 metric (normalized between 0\% and 100\%) was used to quantify distances between parameter vectors. The x- and y-axis tick labels correspond to each other, showing system shape and size, respectively.}
\label{ChapD_Fig4}
\end{figure}

\subsection*{Comparison of extended and original ITWT systems}

Although, by construction, the target patterning behaviors of the extended system and the original ITWT are significantly distinct from each other, we observe some interesting similar features between their posteriors and MAPEs. Note that, for simplicity, we only compare (5, 3, 1) against AI\_L1; here, (5, 3, 1) is the current representative reference, and AI\_L1 is the global reference in \cite{ramirez_sierra_comparing_2025}.

In Fig~\ref{ChapD_Fig5}, we illustrate the comparison between the extended and original ITWT systems. We see that, despite some major differences, the extended and original ITWT posteriors and MAPEs follow a similar trend. For the primary core GRN motif, \textit{Nanog} self-activation (\textit{Nanog}\_NANOG) is lower/stronger than \textit{Gata6} self-activation (\textit{Gata6}\_Gata6), while \textit{Nanog} and \textit{Gata6} mutual repression is relatively balanced (\textit{Gata6}\_NANOG and \textit{Nanog}\_GATA6). For the secondary core GRN motif, \textit{Fgf4} activation by NANOG \hfill\break (\textit{Fgf4}\_NANOG) is significantly higher/weaker than \textit{Fgf4} repression by GATA6 \hfill\break (\textit{Fgf4}\_GATA6), while \textit{Gata6} activation by active ERK (\textit{Gata6}\_A-ERK) is slightly \hfill\break lower/weaker than \textit{Nanog} repression by active ERK (\textit{Nanog}\_A-ERK). We also see that the $\tau_{\textrm{pho},\textrm{NANOG}}$, $\tau_{\textrm{doh},\textrm{NANOG}}$, $\tau_{\textrm{d},\textrm{M-FGFR-FGF4}}$, $\tau_{\textrm{escape}}$, $\tau_{\textrm{exchange}}$, mRNA, and PROTEIN parameter values exhibit strong agreement. However, it is clear that the extended ITWT posterior coverage is broader than the original, especially for the core GRN motif, which suggests the existence of potential compensatory mechanisms and predictive flexibility.

In Fig~\ref{ChapD_Fig6}, we show the conditional model parameter correlation matrix for the extended ITWT, using (5, 3, 1) as our representative reference. Contrary to the AI\_L1 case, we do not see strong linear correlations among the extended ITWT parameter pairs. This lack of a linear parameter-space structure, even for the core GRN interactions, is surprising and hinders our capacity to create hypotheses about compensatory mechanisms among parameter pairs. These highly nonlinear parameter relationships reflect the complex nature of the extended ITWT system, but these results do not shed light on any parameter fine-tuning requirements for the correct blastocyst patterning. For this reason, it is necessary to test the sensitivity to (or robustness against) value perturbations of the inferred model parameters.

In Fig~\ref{ChapD_Fig7}, we show the conditional model parameter sensitivity matrix for the extended ITWT, using (5, 3, 1) as our representative reference. Corroborating our visual inspection, we see low (0-25\%) sensitivities for the extended ITWT parameter singletons and duples. Some clear exceptions to this observation are \textit{Fgf4}\_GATA6, $\tau_{\textrm{d},\textrm{FGF4}}$, $\tau_{\textrm{d},\textrm{M-FGFR-FGF4}}$, and all their respective parameter pairs: the \textit{Fgf4}\_GATA6-related parameters exhibit strong (50-100\%) sensitivities to value changes; the $\tau_{\textrm{d},\textrm{FGF4}}$- and $\tau_{\textrm{d},\textrm{M-FGFR-FGF4}}$-related parameters exhibit moderate (25-50\%) sensitivities to value changes. These sensitivities indicate the importance of these parameter singletons in achieving and sustaining the ideal target system behavior.

\subsection*{Initial luminally deposited FGF4 is critical for supporting the ideal target behavior}

Importantly, the amount of initial luminally deposited FGF4 is a critical player in achieving and sustaining the ideal target system behavior of the extended ITWT. As shown by the box-and-whisker diagram of the FGF4-LUMEN parameter in Fig~\ref{ChapD_Fig3} (bottom-right panel), the extended ITWT system requires a significant amount of initial luminally deposited FGF4 for correctly progressing towards the ideal target behavior: its unconditional posterior interquartile range covers 20\% of its prior range, between 55\% and 75\%, which converts to 7838-10687 FGF4 copies. Notably, this blastocoel-related FGF4 initial condition does not seem to depend on system size or shape. Although this last observation is remarkable, it is also important to consider two underlying aspects. (1) We only examined a few distinct but biologically relevant system size-shape pairs due to computational power restrictions; thus, we cannot discard the idea that a considerably larger ICM population would require a larger amount of initial luminally deposited FGF4. (2) As we characterized in \cite{ramirez-sierra_ai-powered_2024}, the ICM robustly adapts to strongly heterogeneous initial conditions; moreover, the presence of blastocoel and TE provides additional channels for quickly redistributing/degrading any excess FGF4, augmenting the homeostasis capabilities of the system (see Discussion section for complementary details).

Moreover, our analysis for the conditional (on its own MAPE) posterior of the extended ITWT suggests that there are few but interesting parameter relationships involving FGF4-LUMEN. As shown in Fig~\ref{ChapD_Fig6}, there is a moderate (from \textpm~25\% to \textpm~50\%) positive linear correlation between the initial luminally deposited FGF4 and the mean FGF4 escape time from the blastocoel ($\tau_{\mathrm{LUMEN}}$), as well as a moderate negative linear correlation between the initial luminally deposited FGF4 and the FGFR1-FGF4 complex-monomer lifetime ($\tau_{\textrm{d},\textrm{M-FGFR-FGF4}}$). This analysis indicates that both $\tau_{\mathrm{LUMEN}}$ and $\tau_{\textrm{d},\textrm{M-FGFR-FGF4}}$ can concertedly compensate for FGF4-LUMEN parameter value changes, without compromising the ideal target behavior.

As shown in Fig~\ref{ChapD_Fig7}, we also observe that FGF4-LUMEN has an aggregated low (0-25\%) sensitivity to value changes or parameter perturbations. This predictive result is intriguing because it indicates that the unconditional and conditional posterior coverages, with respect to the imposed prior ranges, have a strong agreement. In turn, this relatively large conditional posterior coverage suggests that, although the initial luminally deposited FGF4 is a critical player, it is also a flexible parameter which can cooperatively adjust to system structural changes or perturbations, supporting the ideal target behavior of the extended ITWT.

\begin{adjustwidth}{0in}{0in}
\includegraphics[width = 5.5in, height = 5.5in]{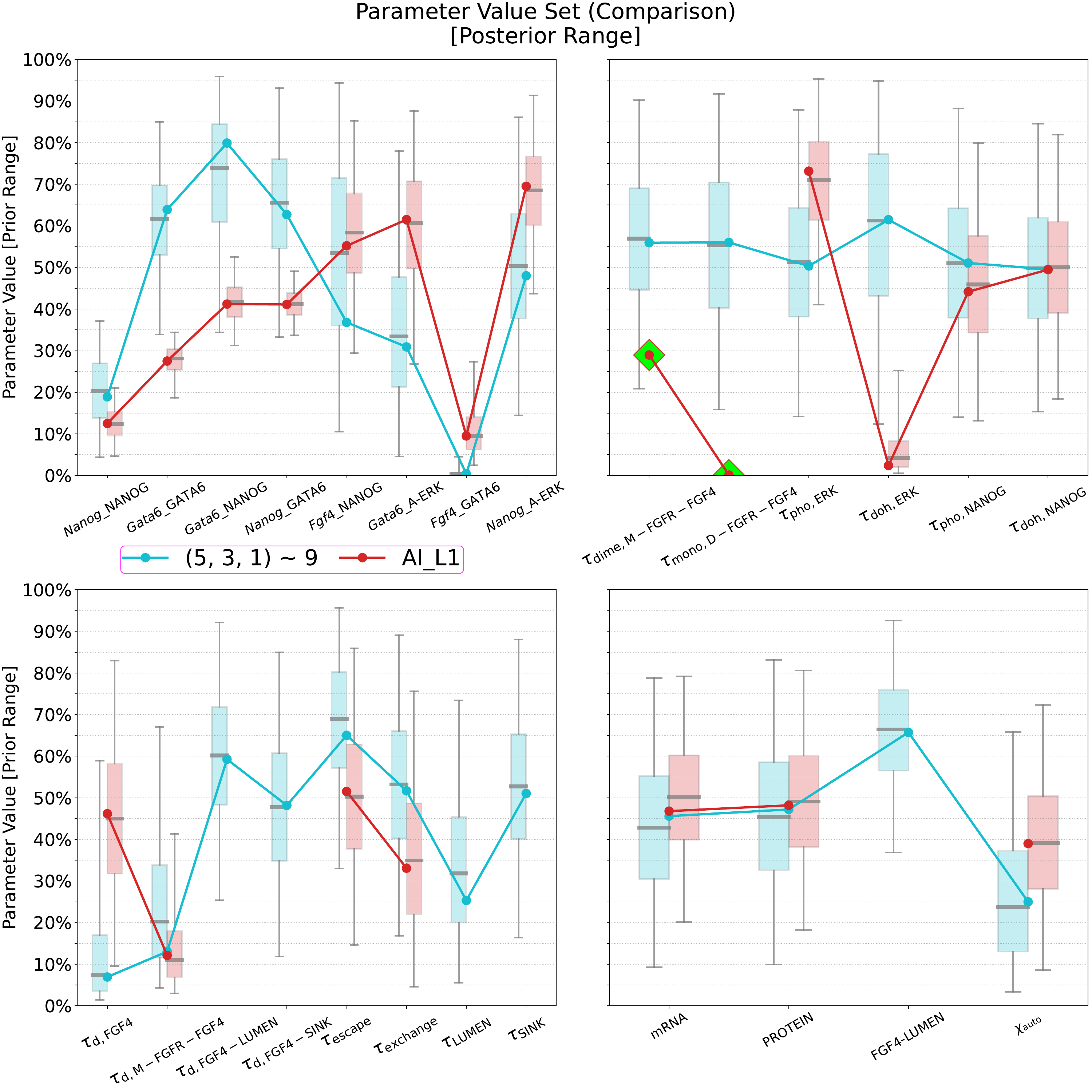} \centering 
\end{adjustwidth}
\begin{figure}[hpt!]
\caption{{\bf Comparison of extended and original ITWT systems.} We compare (5, 3, 1), the current representative reference, against AI\_L1, the global reference in \cite{ramirez_sierra_comparing_2025}. The extended and original ITWT posteriors and MAPEs follow a similar trend. Note that the extended ITWT posterior coverage is broader than the original, especially for the core GRN motif, which suggests the existence of potential compensatory mechanisms and predictive flexibility. The green diamonds highlight the fixed values used for the parameters $\tau_{\textrm{dime},\textrm{M-FGFR-FGF4}}$ and $\tau_{\textrm{mono},\emph{D-FGFR-FGF4}}$ of the original ITWT system; i.e., these two parameter values were not inferred for the original ITWT system.}
\label{ChapD_Fig5}
\end{figure}

\begin{adjustwidth}{0in}{0in}
\includegraphics[width = 5.9in, height = 5.9in]{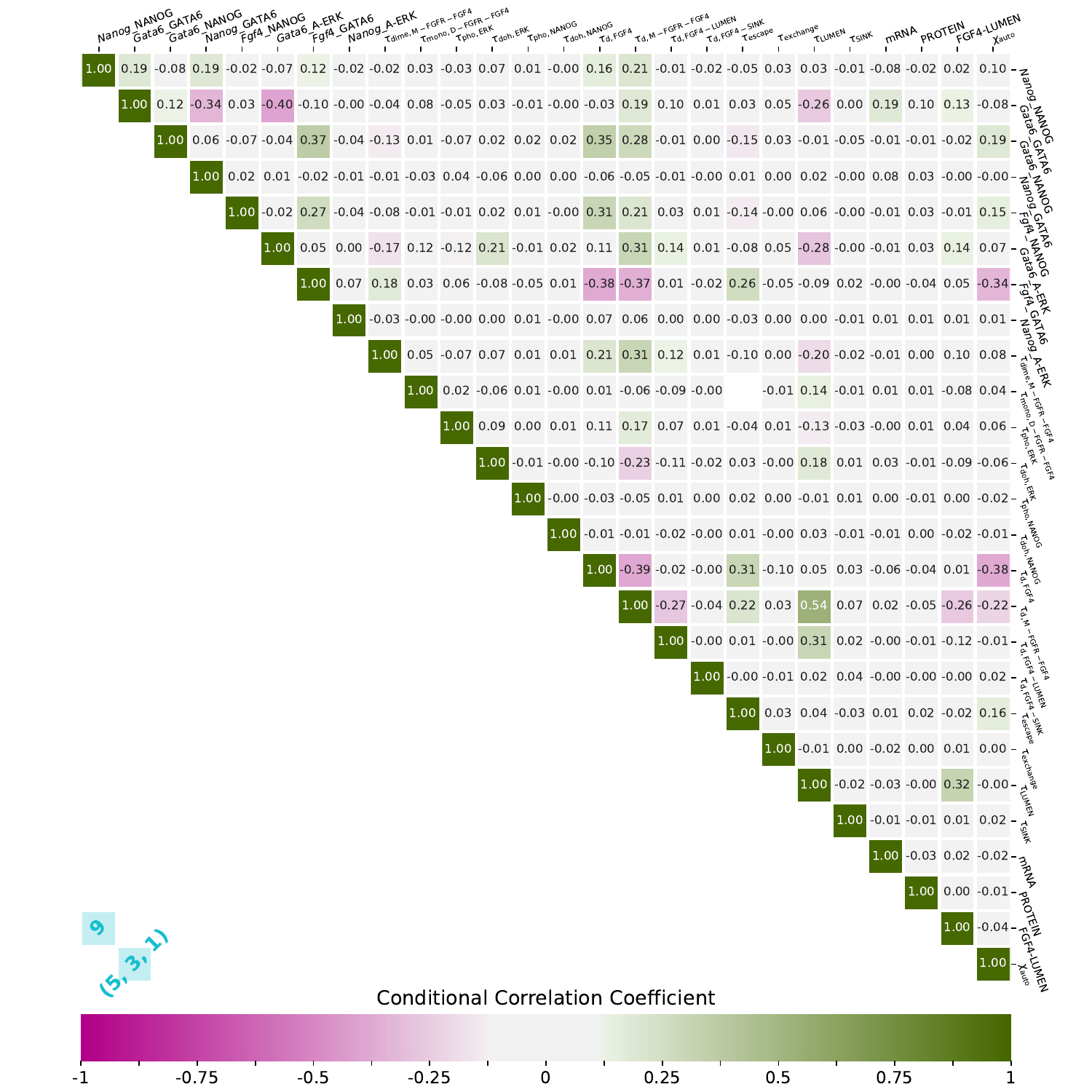} \centering 
\end{adjustwidth}
\begin{figure}[hpt!]
\caption{{\bf Conditional model parameter correlation matrix for the extended ITWT.} Linear correlation coefficients were extracted by conditioning the (5, 3, 1) posterior distribution (representative reference) on its own MAPE. Note that this calculation is possible because the trained ANN acts as a surrogate for the simulator and directly approximates the (5, 3, 1) posterior; i.e., no additional simulations are required to extract Pearson's correlation coefficients.}
\label{ChapD_Fig6}
\end{figure}

\begin{adjustwidth}{0in}{0in}
\includegraphics[width = 5.9in, height = 5.9in]{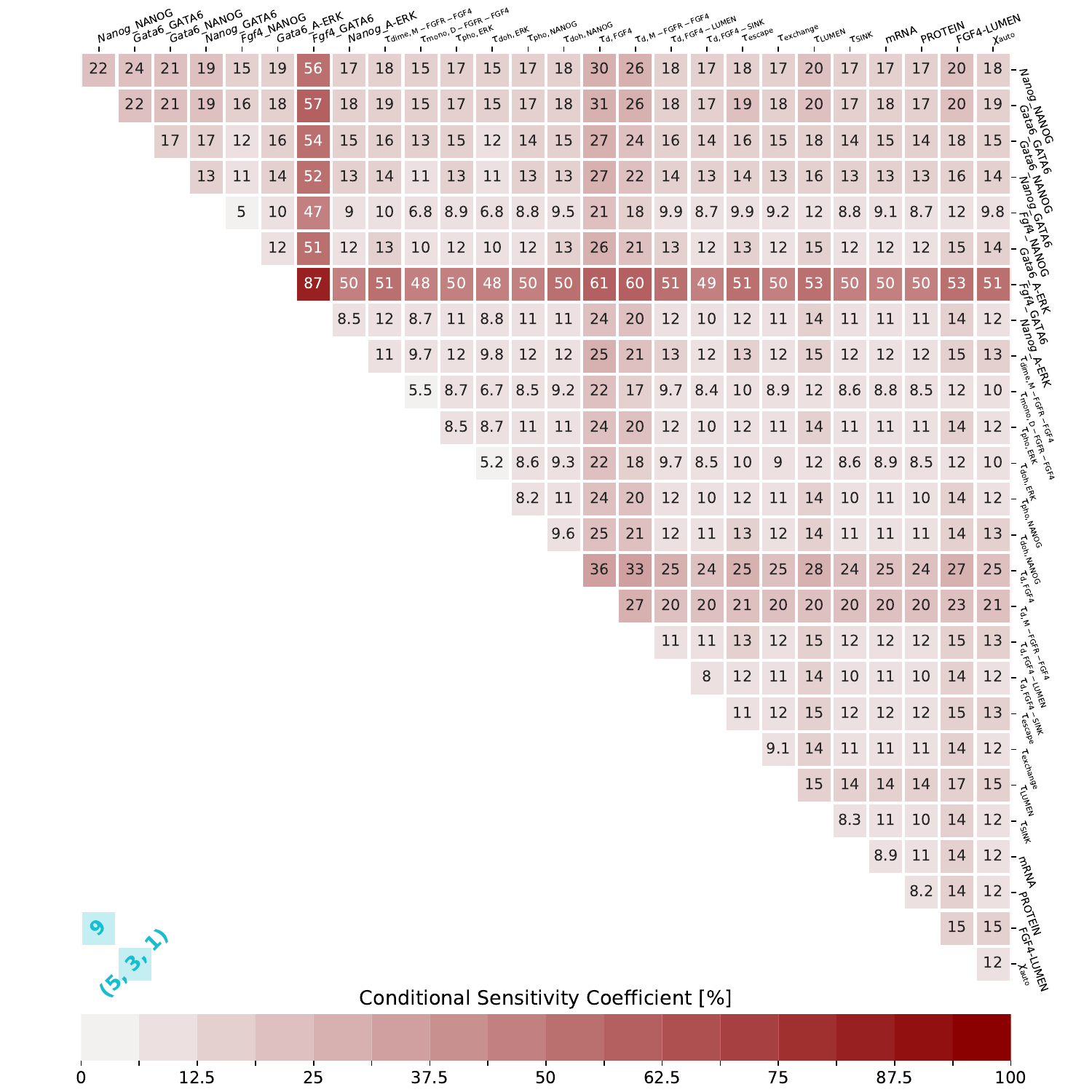} \centering 
\end{adjustwidth}
\begin{figure}[hpt!]
\caption{{\bf Conditional model parameter sensitivity matrix for the extended ITWT.} Sensitivity to value changes for all parameters, with the (5, 3, 1) posterior conditional on its own MAPE. Sensitivity equal to 0\% signifies that any value within its prior range can recover the ideal or target system behavior (while holding all other parameters fixed). Sensitivity equal to 100\% signifies that the value must fall within a singular parameter bin (histograms were created with 250 bins per dimension). Parameter duples (off-diagonal entries) exhibit similar trends to singular (singleton) parameters. Strong sensitivities reflect low tolerance to value fluctuations, given any other parameter is fixed at the corresponding MAPE, for recapitulating the ideal target system behavior.}
\label{ChapD_Fig7}
\end{figure}
\clearpage


\section*{Methods}

In this section, we provide a brief summary of the extended ITWT modeling approach. Interested readers are referred to \cite{ramirez-sierra_ai-powered_2024} for a comprehensive explanation of the computational techniques and fundamental building blocks (bases) of our original ITWT model. Likewise, interested readers are referred to \cite{ramirez_sierra_comparing_2025} for complete details about the posterior correlation and sensitivity analyses, as well as the employed model parameter inference framework (AI-MAPE).

The original ITWT prior has 19 free model parameters. The extended ITWT prior incorporates 7 additional model parameters: two originally fixed parameters, $\tau_{\textrm{dime},\textrm{M-FGFR-FGF4}}$ and $\tau_{\textrm{mono},\emph{D-FGFR-FGF4}}$; two protein lifetimes, $\tau_{\textrm{d},\textrm{FGF4-LUMEN}}$ and $\tau_{\textrm{d},\textrm{FGF4-SINK}}$; two transport parameters, $\tau_{\mathrm{LUMEN}}$ and $\tau_{\mathrm{SINK}}$; one initial condition, FGF4-LUMEN. Thus, the extended ITWT prior features 26 free model parameters in total.

The originally fixed parameters, $\tau_{\textrm{dime},\textrm{M-FGFR-FGF4}}$ and $\tau_{\textrm{mono},\emph{D-FGFR-FGF4}}$, are simply the multiplicative inverse of the nominal rates of FGFR1-FGF4 complex-monomer dimerization and FGFR1-FGF4 complex-dimer monomerization, respectively (see also Fig~\ref{ChapD_Fig5}). The protein lifetimes, $\tau_{\textrm{d},\textrm{FGF4-LUMEN}}$ and $\tau_{\textrm{d},\textrm{FGF4-SINK}}$, are the blastocoel- and TE-associated FGF4 half-lives, respectively. The initial condition, FGF4-LUMEN, is the luminally deposited FGF4 operating as the mean value of a Poissonian-binomial composite distribution, which is sampled at the beginning of every simulation.

The extension of the original ITWT model at the tissue scale includes the aforementioned release parameters, $\tau_{\mathrm{LUMEN}}$ and $\tau_{\mathrm{SINK}}$. These parameters are, respectively, the mean FGF4 escape times from the blastocoel and the TE. In simple terms, any FGF4 molecule, staying or arriving at the blastocoel or the TE, can either degrade or escape towards another part of the blastocyst. To further illustrate this operation, an FGF4 molecule can move omnidirectionally with some minor restrictions (see also Fig~\ref{ChapD_Fig2}): from blastocoel to PRE (bottom cell layer) or TE; from TE to ICM (EPI plus PRE) or blastocoel; from EPI (top cell layer(s)) to PRE or TE; from PRE to EPI or blastocoel or TE.

Moreover, the exit rate of ICM-associated cytoplasmic FGF4 molecules ($1/\tau_{\textrm{escape}}$) gives rise to autocrine and paracrine signaling channels; this FGF4 diffuse flux splitting is controlled via the formula $\chi_{\emph{auto}} + \chi_{\emph{para}} = 1$. The paracrine signaling channel was explicitly adjusted to account for the transport of FGF4 across blastocyst constituents: ICM (EPI plus PRE), blastocoel, and TE. This adjustment is highly technical as it depends on the blastocyst neighborhood architecture (e.g., number of ICM cells and number of cell layers); as such, it was solved algorithmically for each distinct system size-shape pair (see the corresponding simulator scripts available at GitHub \href{https://github.com/MARS-FIAS/}{https://github.com/MARS-FIAS/} for additional details).

\subsection*{Inventory of model parameter values}

Table~\ref{ChapD_Table1} gives an overview of the extended ITWT model parameters designated as free values, providing their respective prior ranges. This table is an extension of the inventory (inferred ``free'' model parameters) presented in \cite{ramirez-sierra_ai-powered_2024} and \cite{ramirez_sierra_comparing_2025}, and it is included here for simplicity; therefore, the original and extended ITWT prior ranges are identical, except for the aforementioned 7 additional model parameters. These additional prior ranges are based on data compiled from various literature sources and generally reflect informed estimates derived from analogous biological systems.

\begin{table}[ht!]
\begin{adjustwidth}{0in}{0in}
\caption{{\bf Summary of inferred (free) model parameters.}}
\adjustbox{max width=\textwidth}{ 
\centering
\begin{tabular}{ | c | c | c | c | c | c | c | }
\hline
\rowcolor{lightgray}
Name & Alias & Description & \parbox[c][2.5em][c]{2.75em}{\centering Lower \\ Bound} & \parbox[c][2.5em][c]{2.75em}{\centering Upper \\ Bound} & Units & Note \\
\hline
$\tau_{\textrm{LUMEN}}$ & $k_{\textrm{LUMEN}}^{-1}$ & \parbox[c][2.5em][c]{8.25em}{\centering Mean escape time \\ (FGF4 Blastocoel)} & 300 & 4500 & [s] & \parbox[c][2.5em][c]{7.5em}{\centering Blastocyst-Wide \\ Communication} \\
$\tau_{\textrm{SINK}}$ & $k_{\textrm{SINK}}^{-1}$ & \parbox[c][2.5em][c]{8.25em}{\centering Mean escape time \\ (FGF4 TE)} & 300 & 4500 & [s] & \parbox[c][2.5em][c]{7.5em}{\centering Blastocyst-Wide \\ Communication} \\
\hline
$\tau_{\textrm{dime},\textrm{M-FGFR-FGF4}}$ & $k_{\textrm{dime},\textrm{M-FGFR-FGF4}}^{-1}$ & \parbox[c][2.5em][c]{8.25em}{\centering Monomer \\ dimerization} & 300 & 345600 & [s] & \parbox[c][2.5em][c]{7.5em}{\centering Signaling \\ Pathway} \\
$\tau_{\textrm{mono},\textrm{D-FGFR-FGF4}}$ & $k_{\textrm{mono},\textrm{D-FGFR-FGF4}}^{-1}$ & \parbox[c][2.5em][c]{8.25em}{\centering Dimer \\ monomerization} & 300 & 345600 & [s] & \parbox[c][2.5em][c]{7.5em}{\centering Signaling \\ Pathway} \\
\hline
\parbox[c][3.75em][c]{7.5em}{\centering Mean Initial \\ FGF4 Count \\ Blastocoel} & \parbox[c][3.75em][c]{7.5em}{\centering FGF4-LUMEN \\ PROTEIN} & Initial condition & 0 & 25000 & [pc] & \parbox[c][2.5em][c]{7.5em}{\centering Signaling \\ Ligand} \\
$\tau_{\textrm{d},\textrm{FGF4-LUMEN}}$ & $k_{\textrm{p},\textrm{d},\textrm{FGF4-LUMEN}}^{-1}$ & \parbox[c][3.75em][c]{8.25em}{\centering Lifetime or half-life \\ Blastocoel} & 300 & 28800 & [s] & \parbox[c][2.5em][c]{7.5em}{\centering Molecular \\ Stability} \\
$\tau_{\textrm{d},\textrm{FGF4-SINK}}$ & $k_{\textrm{p},\textrm{d},\textrm{FGF4-SINK}}^{-1}$ & \parbox[c][3.75em][c]{8.25em}{\centering Lifetime or half-life \\ TE} & 300 & 28800 & [s] & \parbox[c][2.5em][c]{7.5em}{\centering Molecular \\ Stability} \\
\hline
\end{tabular}} 
\begin{flushleft}
Notation: [s] = [seconds]; [pc] = [protein copies].
\end{flushleft}
\label{ChapD_Table1}
\end{adjustwidth}
\end{table}

\newpage


\section*{Discussion}

The crosstalk between gene regulation and morphogenetic change is fundamental during early embryonic development, and within this context, luminogenesis plays an important functional role \cite{dumortier_hydraulic_2019, ryan_lumen_2019, shahbazi_mechanisms_2020}. In particular, during mammalian preimplantation development, the blastocyst lumen influences tissue specification and sorting: within this developmental period, the early layout of the body plan unfolds, and any significant disturbances to this process can lead to embryo fatality \cite{dumortier_hydraulic_2019, ryan_lumen_2019, kim_deciphering_2021}. Therefore, unraveling the elementary mechanisms that coordinate blastocyst cell-fate specification and sorting can help shape our understanding of early pregnancy complications, facilitating the creation of novel therapeutic strategies \cite{kim_deciphering_2021}.

While ethical concerns may limit experimental studies, theoretical and computational models offer a powerful alternative, allowing us to circumvent these limitations. By capturing the complex biophysical dynamics of blastocoel formation and positioning within the mechanical constraining imposed by the TE, these models can reveal novel and critical avenues for research \cite{ryan_lumen_2019, fuji_computational_2022}. Yet, the construction of such models is technically challenging: a comprehensive virtual replica of mammalian blastocyst development will require a mechanistic model that implements not only the stochastic and non-instantaneous nature of gene-expression and cell-cell signaling dynamics, but also the spatial and ever-changing nature of tissue remodeling, morphogenesis, and other embryo-wide processes such as vesicular trafficking.

Here, exploiting the spatial-stochastic simulator and the AI-powered SBI workflow presented in \cite{ramirez-sierra_ai-powered_2024} and \cite{ramirez_sierra_comparing_2025}, we offer a complementary perspective to existing blastocyst morphogenesis models with phenomenological descriptions; i.e., in silico models with primarily deterministic and pseudo-time representations of gene-expression and cell-cell signaling dynamics \cite{tosenberger_computational_2019, cang_multiscale_2021}. Although our current modeling framework bypasses any representation of tissue force interactions and incorporates only minimal morphogenetic features, it enables the simulation of spatially resolved and stochastic developmental trajectories. More importantly, it captures the dynamics of EPI-PRE cell-fate differentiation over biologically relevant timescales, while integrating genuine gene-expression stochasticity (intrinsic noise) and cell-cell signaling delays as essential drivers of nonlinear systematic behavior \cite{eldar_functional_2010, munsky_using_2012}.

Inspired by the experimental study \cite{ryan_lumen_2019}, we posited that our spatial-stochastic simulations would exhibit signatures of the functional roles played by the blastocoel and the TE for canalizing and maintaining the correct blastocyst patterning. Expressly, we hypothesized that the blastocoel functions as a signaling source which accumulates available FGF4, whereas the TE functions as a signaling sink which degrades excess FGF4.

As such, we investigated the synergistic interplay between (cell-level) biochemical stochasticity and (tissue-level) spatial coupling of cell signaling in the presence of a blastocyst lumen and the TE. In contrast to the original ITWT (see \cite{ramirez-sierra_ai-powered_2024}), whose objective was the ideal EPI-PRE cell-fate proportioning, the extended ITWT objective is the ideal blastocyst (spatial) cell-fate patterning: the lowermost cell layer, which is in contact with the blastocoel, must display exclusive commitment to the PRE lineage, and the upper cell layer(s) must display exclusive commitment to the EPI lineage; as characterized by experimental data \cite{bessonnard_icm_2017, hashimoto_epiblast_2019, saiz_growth-factor-mediated_2020, yanagida_cell_2022, allegre_nanog_2022}.

Our findings confirm that, based on the predicted parameter structures and relationships, the blastocoel acts as a localized signaling source and the TE acts as an embryo-wide signaling sink: despite sharing the same prior range, the nominal FGF4 degradation rate in the TE is faster than in the blastocoel, but the nominal mean FGF4 escape rate in the TE is slower than in the blastocoel. This predictive result is nontrivial as no explicit inductive bias was introduced to the the inference procedure for generating the posterior approximations.

However, we also observed that the parameter determining the initial luminally deposited FGF4 strongly influences the progression towards the ideal target system behavior: the extended ITWT requires, on average, 9547 FGF4 copies initially to recapitulate its perfect patterning state. Remarkably, the amount of initial luminally deposited FGF4 does not seem to be dependent on the system size (number of ICM cells) nor shape (number of cell layers), which suggests robustness of the FGF4 signaling mechanism to variations in blastocyst architecture.

This observation contrasts with our previous findings in \cite{ramirez-sierra_ai-powered_2024}, where exogenous FGF4 was shown to finely control EPI-PRE cell-fate proportions, aligning with recent experimental studies \cite{raina_cell-cell_2021, allegre_nanog_2022}. Yet, this observation does not contradict our previous findings because the ICM population is capable of robustly adapting to strongly heterogeneous initial conditions, as also demonstrated in \cite{ramirez-sierra_ai-powered_2024}. For the extended ITWT, the presence of blastocoel and TE enhances the EPI-PRE cellular homeostasis. The blastocoel serves as a highly localized source of initial FGF4, facilitating the canalization of the lowermost cell layer towards the PRE fate, while the TE serves as a surrounding (embryo-wide) sink of surplus FGF4, facilitating the maintenance of the EPI fate across the upper cell layer(s).

On the technical side, given the surprising similarities that we noticed between the predictions for the original and extended ITWT posteriors and MAPEs, it would be interesting to utilize the original ITWT optimal parameter set to investigate the behavior dynamics of the extended system and vice versa. This parameter set swap could potentially boost our mechanistic understanding of how the initial luminally deposited FGF4 dictates the blastocyst-wide cell-fate decision dynamics. This strategy is necessary due to the highly nonlinear parameter-space structure of the extended ITWT, which restricted our analysis of its posterior distributions. We remark that, although it might be counter-intuitive, the inferred posterior distribution of the original ITWT model cannot be directly reused for any other scenario with different prior dimensionality (number of free parameters); overcoming this restriction is an active area of research \cite{vetter_sourcerer_2024, gloeckler_all--one_2024}.

Despite employing a minimal spatial geometry, without cellular divisions or force interactions among cells, this work creates the foundation for a potential multi-scale bridge between our genuine spatial-stochastic representation of single-cell gene expression and intercellular signaling dynamics, with existing agent-based approaches suitable for representing constant tissue-geometry remodeling \cite{andersen_shape_2009, liebisch_cell_2020}. Such an augmented framework would enable the exploration of cell-potency landscapes concomitantly with tissue homeostasis across time and space, capable of simulating highly detailed spatial-stochastic models.

Disentangling the basic mechanisms that drive communication between early lineages, cellular proliferation, blastocoel emergence, and blastocyst patterning is essential for discovering the general factors governing the success of the post-implantation embryo. These processes likely follow conserved principles across mammalian species \cite{dumortier_hydraulic_2019, shahbazi_mechanisms_2020}, making them pivotal for advancing our knowledge of human early development.





\bibliography{Manuscript_Catalog.bib}




\newpage


\section*{Acknowledgments}

The successful completion of this research project owes much to the collaboration and support of esteemed colleagues and collaborators. We extend our deepest gratitude to Sabine Fischer, Tim Liebisch, Franziska Matthäus, and Simon Schardt for their pivotal roles in fostering insightful discussions and providing constructive feedback.

We also express our appreciation to Roberto Covino and his lab for their invaluable guidance and expertise throughout the project, especially for pointing us in the direction of the powerful SBI framework. Their thoughtful insights and constructive critiques have enriched the quality of our work.

Furthermore, we would like to acknowledge the Center for Scientific Computing (CSC) at Goethe University Frankfurt for granting us access to the Goethe-HLR cluster, which has been instrumental in facilitating the progress of this research.

\section*{Funding Information}

This research work was funded by the LOEWE-Schwerpunkt ``Center for Multiscale Modelling in Life Sciences'' (CMMS), which is sponsored by the Hessian Ministry of Science and Research, Arts and Culture{\textemdash}Hessisches Ministerium für Wissenschaft und Forschung, Kunst und Kultur{\textemdash}(HMWK). The funders had no role in study design, data collection and analysis, decision to publish, or preparation of the manuscript.

\section*{Data Availability}

All code files will be available from a GitHub repository. All estimated parameter posterior distributions will be available from the same GitHub repository. The complete data bank will be available from a Zenodo repository.

\section*{Competing Interests}

The authors declare that no competing interests exist.


\end{document}